\documentstyle[12pt,aaspp4]{article}

\lefthead{Bekki et al.}
\righthead{A new model of M32}

\begin{document}
\title{A new formation model for M32: A threshed  early-type spiral?}

\author{Kenji Bekki and Warrick J. Couch} 
\affil{
School of Physics, University of New South Wales, Sydney 2052, Australia}


\author{Michael  J. Drinkwater}
\affil{
School of Physics, University of Melbourne,  Victoria 3010, Australia}

\and
\author{Michael D. Gregg}
\affil{Department of Physics, University of California at Davis, Davis, 
CA 94550. USA}

\begin{abstract}
The origin of the closest  compact elliptical galaxy (cE) M32  
is a  longstanding  problem
of galaxy formation in the Local Group.  
Our N-body/SPH simulations suggest
a new scenario in which  the strong tidal field of M31 can transform  
a spiral galaxy into a compact elliptical.
As a low luminosity spiral galaxy plunges into the central
region of M31,
most of the outer stellar and gaseous components of its disk
are dramatically  stripped due to M31's tidal field.
The central bulge component, on the other hand,
is just weakly influenced by the tidal field
owing to its compact configuration,
and retains its  morphology. 
M31's strong tidal field also induces rapid gas transfer
to the central region, triggers a  nuclear starburst,
and consequently forms the central high density  
and more metal-rich stellar populations with relatively young ages. 
Thus, in this scenario, M32 was previously the bulge of a spiral
tidally interacting with M31
several    Gyr ago. 
Furthermore, we suggest cEs like M32 are rare,  the result of both the rather narrow parameter space for tidal interactions which morphologically transform spirals into cEs
and the very short time scale ($<$ a few $10^9$ yr) for cEs to be swallowed
by their giant host galaxies (via dynamical friction) after their  formation. 
\end{abstract}


\section{Introduction}
Andromeda's (M31) closest companion, M32, has long served as
a laboratory that can provide valuable information  
not only on the nature of stellar populations of low mass elliptical galaxies
but also on the detailed structure of their nuclei (e.g., Davidge 2000; van den Bergh 2000).
M32 is classified, morphologically, as a compact elliptical (cE) 
that has rather high central  surface brightness 
($\sim$ four orders of magnitude higher than that
of typical sheroidals of comparable total luminosity),
a truncated de Vaucouleurs profile, 
and solar-like metallicity (e.g., Kormendy 1985; Nieto \& Prugniel 1987).
M32-like cEs are observed to be located almost exclusively in the vicinity
of bright galaxies (Nieto \& Prugniel 1987) and generally considered to be very
rare objects both in the field and in clusters
(Kormendy 1985; Ziegler \& Bender 1998; Drinkwater \& Gregg 1998).
A large number of previous spectroscopic studies aimed at placing strong constraints
on M32's star formation history have  suggested the existence of 
young stellar populations with ages of several Gyr (e.g., O'Connell 1980; Burstein  et al. 1984; Rose 1984;
Bica, Alloin, \& Schmidt 1990; Vazdekis \& Arimoto 1999;
del Burgo et al. 2001), 
though recent studies have argued that there is no clear evidence
supporting M32's relatively recent star formation (e.g., Cole et al. 1998; Renzini 1998).
The most recent spectroscopic studies have suggested
M32 has relatively young ($\sim$ several Gyr old) stellar populations 
(e.g., del Burgo et al. 2001; Davidge et al. 2001; 
See  Davidge 2000 and van den Bergh 2000 for a  recent review).


The origin of the  high surface brightness of cEs and radially-limited luminosity profiles
has been discussed mostly in the context of
the tidal effects of massive galaxies close to them 
(King 1962; Faber 1973; Nieto \& Prugniel 1987; Burkert 1994). 
For example, Faber (1973) first proposed  that M32 was previously
the inner high-density  region of a low-luminosity elliptical galaxy 
with the outer part stripped by M31 tidal field. 
Burkert (1994) proposed an alternative scenario that
M32 formed through starbursts
and the subsequent violent gravitational collapse in M31's  vicinity,
where its strong tidal field  can induce  very efficient violent relaxation
in the M32 collapse.
It is, however, still controversial how these models can provide a plausible explanation 
for the observationally suggested intermediate-age  ($\sim$ 5 Gyr) stellar populations
in M32.

Through numerical simulations, we have investigated how M31's
strong tidal field affects the chemical and dynamical evolution
of a gas-rich spiral plunging into the central region of M31,
demonstrating that dissipative tidal interactions can dramatically
transform a spiral into a cE such as M32.
In this scenario,  a gas-rich disk galaxy with a compact
bulge, captured by M31 several Gyr ago, loses a significant fraction of its initial
disk mass during dynamical interaction with M31 through tidal stripping.
The central bulge component, on the other hand,
survives,  retaining its compactness. 
Although this ``threshed spiral'' (for the concept of galaxy ``threshing'', see
Bekki, Couch, \& Drinkwater 2001; hereafter BCD) scenario
has already  been speculated  on by several authors (e.g., Nieto 1990; van den Bergh 2000),
the present numerical study is the first to demonstrate that this scenario
is realistic and viable for the formation of M32.
This scenario is in a striking contrast to the canonical   one (Wirth \& Gallagher 1984;
Kormendy 1985)
in which cEs like M32 form the faint branch of the normal elliptical sequence.


\section{Model}

We consider a gas-rich disk galaxy with a bulge,  
orbiting a massive disk galaxy with structural  and kinematical properties
similar to those of M31.  
We adopt TREESPH codes described in  Bekki (1995) for hydrodynamical evolution
of galaxies.
We use the disk model of Fall-Efstathiou (1980) with a dark-halo-to-disk mass
ratio equal to 4 for M32's progenitor disk.
The total mass ($M_{\rm d}$) and  the size of the exponential disk ($R_{\rm d}$) are 
4.0 $\times$ $10^9$ $M_{\odot}$ and 4.5 kpc (scale length of 0.9 kpc),  respectively. 
The gas mass fraction is set to be 0.1 
and an  isothermal equation of state is used for the gas 
with a temperature of $7.3\times 10^3$ K, corresponding to a sound speed 
of 10 km $\rm s^{-1}$.
Star formation is modeled 
according to the Schmidt law (Schmidt 1959)
with an exponent of 1.5.
The central bulge is modeled by the Plummer model (Binney \& Tremaine 1987)
with total mass 2.0 $\times$ $10^9$ $M_{\odot}$, corresponding to
a bulge mass fraction of 0.33 and scale length of 0.25 kpc.
The total number of particles used for each model 
are 20,000 for collisionless particles and 5,000 for gaseous ones.

The orbit of the  spiral  is assumed to be influenced
only by the {\it fixed} gravitational potential of M31,
having three components: a dark matter halo, a disk,
and a bulge. We assume  a logarithmic dark matter halo potential,
\begin{equation}
{\Phi}_{\rm halo}=v_{\rm halo}^2 \ln (r^2+d^2),
\end{equation}
a Miyamoto-Nagai (1975) disk,
\begin{equation}
{\Phi}_{\rm disk}=-\frac{GM_{\rm disk}}{\sqrt{R^2 +{(a+\sqrt{z^2+b^2})}^2}}
\end{equation}
and a spherical Hernquist (1990) bulge 
\begin{equation}
{\Phi}_{\rm bulge}=-\frac{GM_{\rm bulge}}{r+c},
\end{equation}
where $r$ is the distance from the center of M31,
$d$ = 12 kpc, $v_{\rm halo}$ = 131.5 km ${\rm s}^{-1}$,
$M_{\rm disk}$ = 1.3 $\times$ $10^{11}$ $M_{\odot}$,
$a$ = 6.5 kpc, $b$ = 0.26 kpc, $M_{\rm bulge}$ =  9.2 $\times$ $10^{10}$ $M_{\odot}$,  
and $c$ = 0.7 kpc. This reasonable set of parameters gives a realistic 
rotation curve for M31 with a rotation speed of 260 km ${\rm s}^{-1}$ at 26 kpc.

The center of M31 is always
set to be ($x$,$y$,$z$) = (0,0,0) whereas the initial position
and velocity of the spiral  are ($x$,$y$,$z$) = (0, 0,$r_{\rm in}$) 
and ($V_{\rm x}$,$V_{\rm y}$,$V_{\rm z}$) = (0,$V_{\rm in}$,0),
respectively. By changing these two parameters
$r_{\rm in}$ and $V_{\rm in}$, we investigate how the
transformation process from spirals  into  cEs  depends  on their orbits.  
Although we have investigated models 
with a variety of different $r_{\rm in}$ and $V_{\rm in}$,
we mainly describe here the results of a `standard' model
with $r_{\rm in}$  = 12 kpc and $V_{\rm in}$ = 142 km $\rm s^{-1}$ 
(corresponding to $0.5V_{\rm c}$,
where $V_{\rm c}$ is a circular velocity of M31 at this radius). 
Figure 1 shows the orbit  with respect to M31
and  the final mass distribution for 
the simulated M32 in the standard model. We choose this nearly polar orbit
from  the most likely M32 orbit by Cepa \& Beckman (1988).

In the following, our units of mass, length, and time are
2.0 $\times$ $10^8$ M$_{\odot}$ 
7.94 $\times$ $10^2$ pc, 
and 2.36 $\times$ $10^7$ yr, respectively. 
Parameter values  and final morphologies for each model
are summarized in Table 1. The fourth  column gives
pericenter distance ($r_{\rm p}$)
for each model.
The fifth describes the final morphological properties  after 32 time units 
(corresponding to 0.76 Gyr): 
`cE' indicates a remnant with the outer disk completely stripped yet 
the bulge  largely unaffected, `S0' the case where both the 
outer disk and bulge  survive yet the disk is partly
stripped (thus becoming a smaller disk) and rather thickened
owing to dynamical heating,  `no remnant' where both disk and bulge components
are tidally destroyed, and 'Sa' the case
where the spiral keeps its initial morphology.

\begin{figure}
\epsscale{1.0}
\plotone{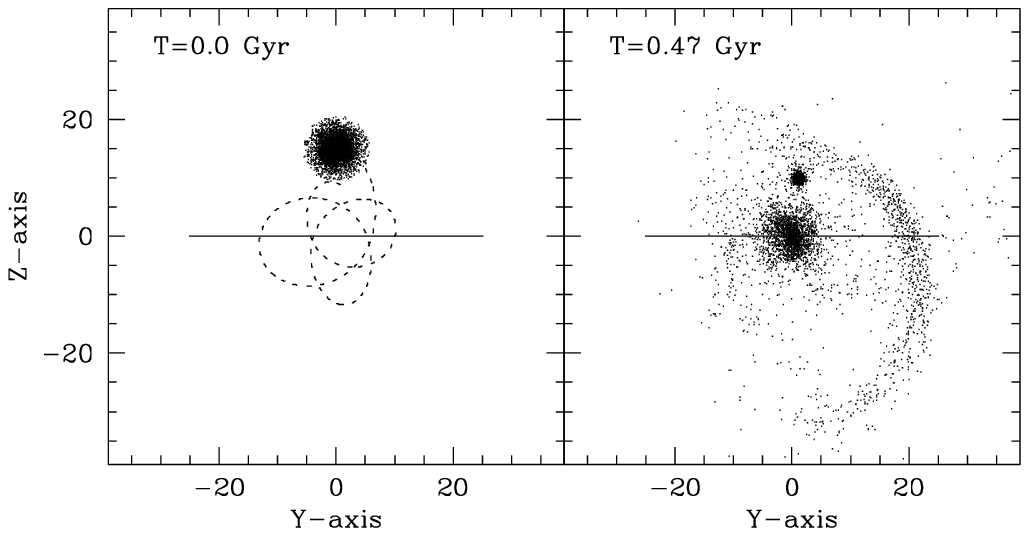}
\caption{
Initial ({\it left}) and final ({\it right})  mass distribution of the simulated spiral projected onto the $y$-$z$ (face-on) plane for the ``standard'' model. 
The orbital evolution of  the  spiral with respect to
the center of M31 is also given by {\it dashed} line in the left panel
in order that the dynamical evolution can be seen more clearly. 
The center of M31 is  set to be always ($x$,$y$,$z$) = (0,0,0) and the
initial and final positions  of the spiral are  ($x$,$y$,$z$) = (1.3,0,15.0)
and (1.0,1.3,8.5), respectively. 
The scale is given in our units (0.8 kpc) and thus
each frame measures 62 kpc on a side. The initial 
position and size of M31's disk are  given by a {\it solid} line
for each panel.
Note that owing to the strong tidal field of M31,
a significant fraction of the outer stellar disk
of the spiral (both old and young stars) is stripped away at $T$ = 0.47 Gyr:
only a compact bulge can be seen at ($x$,$y$,$z$) = (1.0,1.3,8.5)
(The apparently high surface density clumping of stars around  ($x$,$y$,$z$) = (0,0,0)
is due to tidal debris of the spiral).
Note also that without including the effects of dynamical friction
in the model, the apocenter distance of the orbit becomes smaller 
owing to the mass loss of the spiral preferentially at its pericenter.
The surface brightness of the tidal debris is estimated to 
be $\sim$ 26 mag ${\rm arcsec}^{-2}$ in $V$-band for the M32's distance
of 760 kpc from the Galaxy. 
\label{fig-1}}
\end{figure}

\begin{figure}
\epsscale{0.6}
\plotone{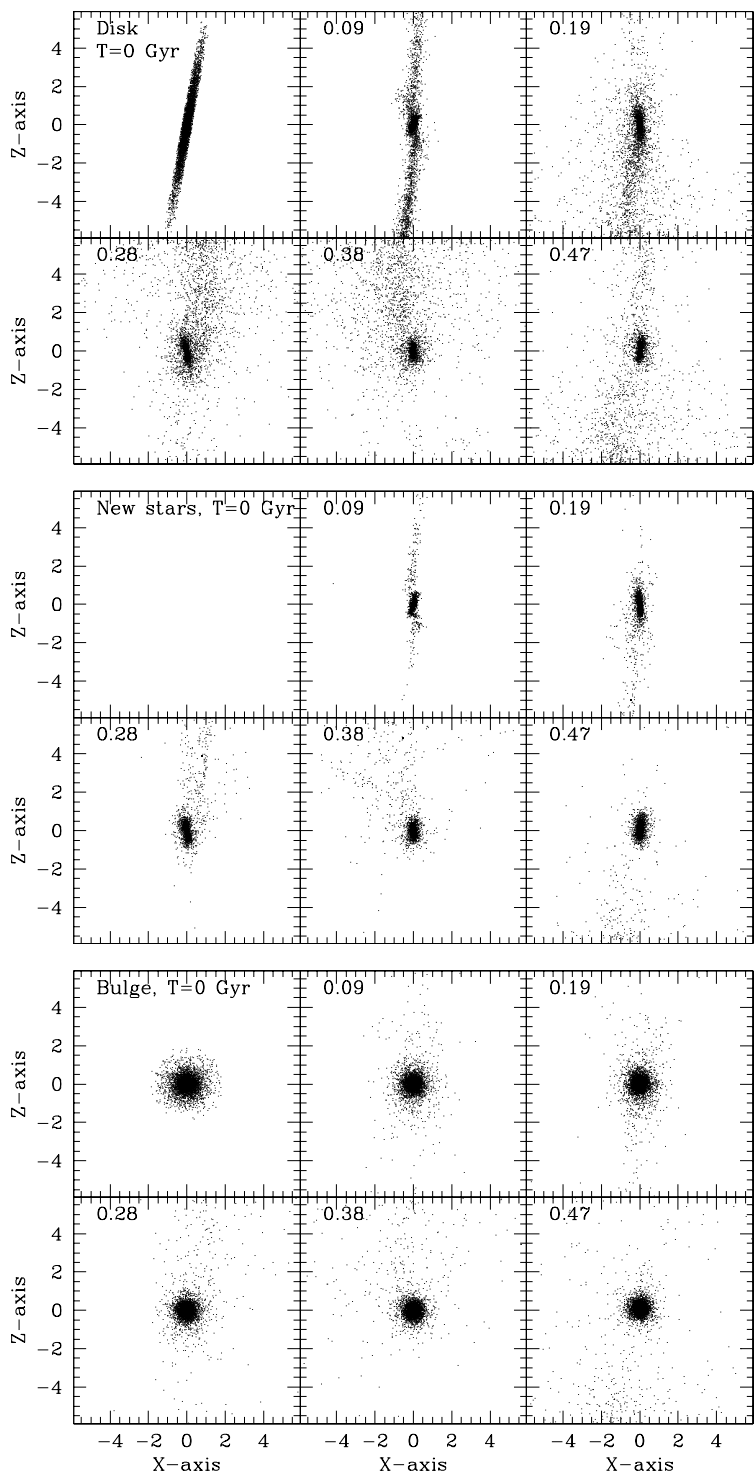}
\caption{
Morphological evolution projected onto the $x$-$z$ plane (edge-on)
for the stellar disk (top six panels),
the new stars formed from gas (middle),
and the bulge (bottom) 
in the simulated spiral. 
The time indicated in the upper left
corner of each frame is given in Gyr and
each frame measures 9.4 kpc on a side.
\label{fig-2}}
\end{figure}

\placefigure{fig-1}
\placefigure{fig-2}

\section{Results}

As the spiral  approaches the pericenter of its orbit for the first time 
($T$ = 0.09 Gyr), the strong tidal field of M31  stretches the stellar disk 
along the direction of the spiral's  orbit and consequently tidally strips
the stars of the disk  (see Figure 2).
The disk gradually loses a large number of its stars  from its
outer part (0.09 $<$ $T$ $<$ 0.38 Gyr) and finally loses most of its outer stellar 
disk ($T$ = 0.47 Gyr): 
The spiral  can keep only stars initially in the central region where
the bulge gravitational field dominates.
Although galaxy-scale star formation is triggered by the tidal interaction,
the outer new stellar component is also tidally stripped away from the spiral,
and consequently only the central starburst component survives.
The central bulge, on the other hand, is just weakly influenced by the 
tidal force because of  its compact configuration during the tidal destruction
of the outer spiral.

Figure 3 shows that a significant fraction ($\sim$ 59\%) of the stellar disk
is stripped by the M31 strong tidal field within  0.75 Gyr. 
Thanks to its strongly self-gravitating nature,
the bulge loses only a small amount ($\sim$ 19 \%) of mass and thus
can keep its compact morphology during its tidal interaction with M31.
The total mass of new stars within the central 5 kpc suddenly becomes
larger early in the tidal interaction ($T$ $<$ 0.2 Gyr)
because rapid gaseous inflow to the central region causes efficient star formation.
However the mass fraction of the stellar disk and that of the bulge 
do not so significantly change within the bulge-dominated region ($r$ $<$ 2kpc):
The fractional mass decreases (increases) from 0.55 (0.45) to 0.33 (0.59) 
for the disk (bulge).
The fractional mass of the new stellar component, on the other hand,
greatly increases  to reach 0.08, which implies that the simulated M32  can have
a remarkable fraction of young stars 
for an elliptical galaxy. 

The tidal distortion forms non-axisymmetric structures
(spirals and bar), induces  efficient gaseous dissipation in the shocked
gaseous region with the subsequent rapid radial gas transfer to the central region,
which consequently triggers  a massive starburst. This starburst trigger  mechanism is essentially the same as that already demonstrated 
by Noguchi \& Ishibashi (1986) for the case
of a galaxy-galaxy interaction. 
The maximum star formation rate is 9.5 $M_{\odot}$ ${\rm yr}^{-1}$
(at $T$=0.07 Gyr)  corresponding to $\sim$ 31 times higher than the mean star formation rate of 
the  isolated disk model (0.31 $M_{\odot}$ ${\rm yr}^{-1}$).
Owing to this starburst, a significant fraction  of gas (64 \% corresponding to 2.2 times
larger than the isolated model case) is consumed by the star formation,
and most of the remaining gas is tidally stripped away from the disk.
Owing to chemical evolution associated with this efficient star formation,
the bulge finally has a young stellar population with a mean metallicity
of [Fe/H]=$+0.1$ and a radial metallicity gradient.
The central, more metal-rich components and the  gradient may well be responsible for M32 being observed to have stronger CN and Mg absorption for its luminosity (Faber 1973)
and partly associated with 
the origin of the radial variations of total AGB luminosity 
observed in M32 (Davidge et al. 2001).

The present parameter study also provides a clue to the problem
as to why cEs like M32 are apparently so rare (see Table 1). 
For Model 3, with smaller $r_{\rm in}$ (12 kpc)
and smaller $V_{\rm in}$ (28  km $\rm s^{-1}$),
both the bulge and stellar disk components completely
disintegrate during the tidal interaction and, accordingly, 
no remnant is left. 
On the other hand, for Model 5 with  smaller $r_{\rm in}$ (12 kpc) 
and larger $V_{\rm in}$ (255 km $\rm s^{-1}$),
the tidal field can only  heat up the stellar disk in the vertical direction
and consequently transform  a spiral into an S0  with a thick disk. 
These results imply that both  $V_{\rm in}$ and $r_{\rm in}$
should be within a certain range and thus that the 
pericenter, $r_{\rm p}$, of the orbit of a spiral
plunging into M31 should be 
within a certain range ($r_{\rm p}$ $\sim$ several  kpc).
The results of Models 6, 7, and 8 confirm this,
and accordingly the present results strongly suggest
that the parameter space for the formation of cEs like M32
is rather narrow. 
Furthermore, our results demonstrate that S0s can be formed
from spirals with bulges owing to   tidal interaction
with their bright host galaxies (see Models 5 and 7).
 
\begin{figure}
\epsscale{1.0}
\plotone{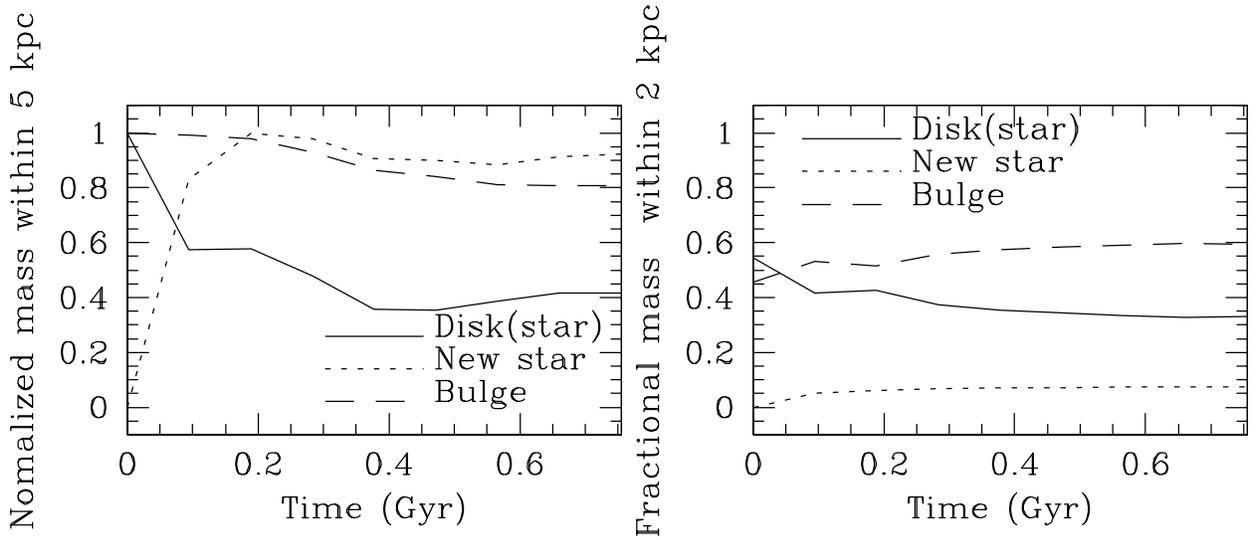}
\caption{
{\it Left}: Time evolution of normalized mass within
the central 5 kpc (corresponding to the disk size of the spiral) for the stellar disk (solid),
the new stars formed from gas (dotted), and the bulge (short-dashed)
in the simulated spiral. Here the mass of the stellar disk and that of the bulge
are normalized by their initial masses  
whereas that of the new stars is normalized by its maximum mass within the 5 kpc.
{\it Right}: Time evolution of fractional  mass within the central 2 kpc
(corresponding  to $\sim 2\times$ larger than the initial size of the bulge)
for the stellar disk ({\it solid}), the new stars  ({\it dotted}), and the bulge ({\it short-dashed}).
\label{fig-3}}
\end{figure}

\begin{figure}
\epsscale{1.0}
\plotone{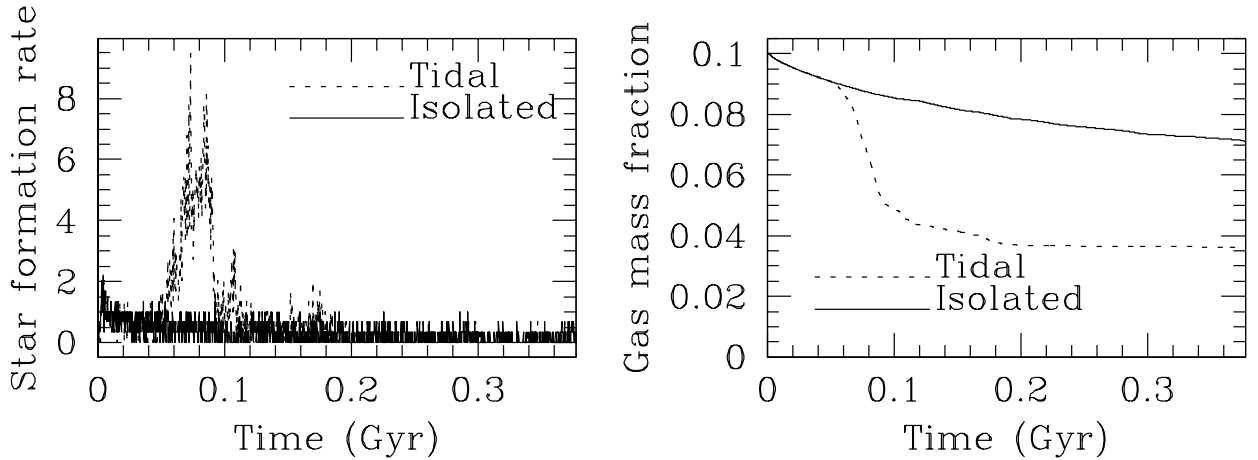}
\caption{
The time evolution of star formation in units of $M_{\odot}$ ${\rm yr}^{-1}$ 
({\it left}) and that of gas mass fraction ({\it right}) for the isolated model ({\it solid}) and for the standard model ({\it dotted}, represented as ``tidal'' model). 
\label{fig-4}}
\end{figure}

\placefigure{fig-3}
\placefigure{fig-4}

\section{Discussion and conclusion}

A new type of sub-luminous and extremely  compact ``dwarf galaxy'' 
(the so-called ``ultra-compact dwarf'', referred to as UCD)  has
been recently discovered in an ``all-object''  spectroscopic survey
centred on the Fornax Cluster (Drinkwater et al. 2000a, b).
Drinkwater \& Gregg (1998),  however, did not find any
promising candidates of cEs (such as NGC 4486B observed in
the vicinity of M87) in the same cluster environment. 
These recent observations have raised the following question
on the origin of two different types of compact galaxy, UCDs and cEs
(Drinkwater et al. 2000a; BCD): Are there any physical
relationships between these two types? 
BCD have demonstrated that the strong tidal field
of a bright elliptical can transform a nucleated dwarf (dEn) into an UCD owing to the  
dramatic stripping of the outer stellar envelope: 
Only the central compact nucleus can manage to survive after
this ``galaxy threshing''. 
This result combined with the present numerical results 
therefore suggests that the  origin of the difference in physical properties
between UCDs and cEs is due to the difference
in morphological types (i.e., spirals versus dEn) 
of the precursor galaxies destined to suffer from ``galaxy threshing''. 
This explanation is consistent with the observational fact
that both UCDs and cEs are located  
preferentially in the vicinity of bright galaxies, 
where ``galaxy threshing'' works more readily. 
Although no UCD's are currently known around M31, there may be one in
its future.  Zinnecker \& Cannon (1986) recognized that the
star-forming dwarf companion of M31 NGC~205 is comparable to Virgo
cluster nucleated dE galaxies; if its orbit with respect to
M31 is rather eccentric, it
will be threshed by M31 and leave an UCD or an $\omega$Centauri-like peculiar
globular cluster (e.g., G1) with $M_{\rm B}$  approximately  $-12$mag around M31.


A direct observational test  for confirming  
the proposed ``threshed spiral'' scenario 
is to investigate whether (1)\,some of M31's halo globular clusters (GCs) 
show peculiarities in their spatial distribution and kinematics,
such as clumping or streams along M32's  orbit  
and retrograde orbital rotation (with respect to M31)
similar or  nearly identical to  M32's retrograde orbit suggested by Cepa \& Beckman (1988),  
and (2)\,there are any stellar and/or gaseous
streams or sub-structures {\it well outside} the present orbital plane of M32.
M32 is observed to have no GCs (Harris 1991),
whereas it should have $\sim$ 10-20 given its luminosity (van den Bergh 2000). 
This fact probably implies that most of the GCs initially associated with M32's  bulge are tidally stripped from M32
at the epoch of M32 formation (e.g., van den Bergh 2000).
Accordingly, if future  photometric and spectroscopic studies, deriving $both$ 
the detailed two-dimensional GC distribution and the radial velocity of each GC  with respect to M31, 
find some GCs located previously on M32's orbit,
these would strengthen the possibility of the threshed scenario.
Furthermore, the present study predicts that tidally stripped stars
can be distributed well outside M32's present orbit.
Such low surface brightness stellar components (e.g.,
the so-called ``spaghetti'' structures observed in the Galaxy;
Morrison et al. 2000) {\it with old and intermediate ages} 
may well be observed
as sub-structures in M31's stellar halo if they are projected onto the sky.
Therefore, future large systematic surveys of M31's stellar halo 
light distribution via wide-field CCD imagery on 8m-class telescopes
(e.g., SUBARU with Suprime-Cam) 
will assess the validity of the proposed ``threshed spiral'' scenario. 

\acknowledgments

We are  grateful to the anonymous referee for valuable comments,
which contribute to improve the present paper.
The authors acknowledge 
the financial support of the Australian Research Council 
throughout the course of this work.
MDG acknowledges support by the National Science Foundation
under Grant No.~9970884.

\clearpage


\clearpage

\begin{deluxetable}{cccccc}
\footnotesize
\tablecaption{Results of different models of tidal interaction between
a spiral galaxy  and  M31 \label{tbl-1}}
\tablewidth{0pt}
\tablehead{
\colhead{model number} 
& \colhead{$r_{\rm in}$ (kpc)} 
& \colhead{$V_{\rm in}$ (km ${\rm s}^{-1}$)} 
& \colhead{$r_{\rm p}$ (kpc)} 
& \colhead{final morphology}
& \colhead{comments}}
\startdata
1 & 12 & 142 & 4.5  & cE & standard model \\
2 & -- & --  & --   & Sa & isolated  model \\
3 & 12 & 28  & 0.5  & no remnant& \\
4 & 12 & 212  & 8.8  & cE &\\
5 & 12 & 255 & 10.7 & S0 & \\
6 & 24 & 50  & 2.2  & cE &  \\
7 & 24 & 188 & 15.0 & S0 & \\
8 & 24 & 376 & 24.0 & Sa & circular orbit\\
\enddata

\end{deluxetable}

\end{document}